\documentclass[aps,prb,reprint,superscriptaddress,amsmath,amssymb,floatfix]{revtex4-2}
\usepackage[T1]{fontenc}
\usepackage[utf8]{inputenc}
\usepackage{graphicx}
\usepackage{textcomp}
\usepackage{microtype}
\usepackage{xcolor}
\usepackage{hyperref}
\hypersetup{colorlinks=true,linkcolor=black,citecolor=blue,urlcolor=blue}
\graphicspath{{figures/}}

\begin{document}
\title{Field-free superconducting diode and tunable vortex dynamics in NbN/FePt stripe-domain heterostructures}

\author{Juan C. Zapata}
\affiliation{Instituto de Nanociencia y Nanotecnolog\'{i}a (CNEA--CONICET), Nodo Bariloche, Av. Bustillo 9500, 8400 San Carlos de Bariloche, Argentina}
\affiliation{Instituto Balseiro, Universidad Nacional de Cuyo and Comisi\'{o}n Nacional de Energ\'{i}a At\'{o}mica, Av. Bustillo 9500, 8400 San Carlos de Bariloche, Argentina}
\author{Emilio De Biasi}
\affiliation{Instituto de Nanociencia y Nanotecnolog\'{i}a (CNEA--CONICET), Nodo Bariloche, Av. Bustillo 9500, 8400 San Carlos de Bariloche, Argentina}
\affiliation{Instituto Balseiro, Universidad Nacional de Cuyo and Comisi\'{o}n Nacional de Energ\'{i}a At\'{o}mica, Av. Bustillo 9500, 8400 San Carlos de Bariloche, Argentina}
\author{Martin Sirena}
\affiliation{Instituto de Nanociencia y Nanotecnolog\'{i}a (CNEA--CONICET), Nodo Bariloche, Av. Bustillo 9500, 8400 San Carlos de Bariloche, Argentina}
\affiliation{Instituto Balseiro, Universidad Nacional de Cuyo and Comisi\'{o}n Nacional de Energ\'{i}a At\'{o}mica, Av. Bustillo 9500, 8400 San Carlos de Bariloche, Argentina}
\author{Jeehoon Kim}
\email{jeehoon@postech.ac.kr}
\affiliation{Department of Physics, Pohang University of Science and Technology, Pohang 37673, South Korea}
\author{Nestor Haberkorn}
\email{nestor.haberkorn@ib.edu.ar}
\affiliation{Instituto de Nanociencia y Nanotecnolog\'{i}a (CNEA--CONICET), Nodo Bariloche, Av. Bustillo 9500, 8400 San Carlos de Bariloche, Argentina}
\affiliation{Instituto Balseiro, Universidad Nacional de Cuyo and Comisi\'{o}n Nacional de Energ\'{i}a At\'{o}mica, Av. Bustillo 9500, 8400 San Carlos de Bariloche, Argentina}

\begin{abstract}
We demonstrate reconfigurable, field-free superconducting diode behavior in NbN/FePt and NbN/AlN/FePt heterostructures with remanent stripe domains in the FePt layer. NbN films 10 nm thick are capped either directly with FePt or through a thin AlN spacer, which reduces electronic proximity effects while preserving magnetic coupling through stray fields. At zero applied magnetic field, the superconducting diode efficiency reaches approximately 50--60\% in NbN/FePt when the stripes are perpendicular to the current and is strongly reduced when they are parallel. The diode effect persists in NbN/AlN/FePt with efficiencies reaching approximately 35--40\%, supporting a magnetic origin of the nonreciprocity. We interpret the response as asymmetric vortex transport in the remanent magnetic landscape, with preferential edge vorticity and a Lorentz force that favors inward flux injection or outward motion depending on current polarity. Under applied perpendicular magnetic fields, the interplay between the external field and the magnetic stray fields modifies the flux-flow instability and the apparent instability velocity inferred from transport, with values below those reported for bare NbN films. These results show that stripe-domain ferromagnetic overlayers provide a reconfigurable route to control superconducting nonreciprocity and flux-flow stability in NbN-based heterostructures.
\end{abstract}

\maketitle

\section{Introduction}

The study of vortex dynamics in micro- and nanoscale superconducting systems is essential for both fundamental research and cryogenic device technologies \cite{ref1,ref2,ref3}. The performance of these devices is governed by the interplay between geometric dimensions and intrinsic superconducting parameters such as the critical temperature ($T_c$), penetration depth (\ensuremath{\lambda}), and coherence length (\ensuremath{\xi}). Surface barriers and vortex motion, which ultimately set the onset of dissipation, play a central role in advanced devices such as superconducting nanowire single-photon detectors (SNSPDs) \cite{ref4,ref5} and superconducting diodes \cite{ref6}. In both device classes, vortex entry and subsequent vortex dynamics can significantly influence the onset of dissipation under bias, especially near the critical current and under applied magnetic fields. In SNSPDs, a dc bias current close to the critical current enables the detection of single photons through resistive transitions. While operation is conventionally described in terms of hot-spot formation, under applied fields the hot spot can also facilitate thermally activated vortex crossing and depinning, thereby enhancing detection efficiency. Controlling vortex penetration and dynamics is therefore key for optimizing sensitivity, speed, and noise performance in these devices \cite{ref5,ref7,ref8,ref9}. Likewise, superconducting diodes can exploit asymmetric barriers for flux injection and vortex motion to produce nonreciprocal current--voltage (I--V) characteristics and a polarity-dependent onset of dissipation \cite{ref3,ref6,ref10}.

Beyond geometry and intrinsic properties, the functionality of superconducting devices can be tailored through interfacial proximity effects. In superconductor/ferromagnet (S/FM) hybrids, both electronic proximity coupling and magnetic stray fields (MSF) originating from the FM can strongly modulate the superconducting response \cite{ref11}. Such interactions can shift $T_c$ \cite{ref12}, improve detector recovery times \cite{ref13}, or create asymmetric surface barriers for superconducting diodes \cite{ref14,ref15,ref16,ref17,ref18}. Interface transparency controls the electronic proximity contribution, while the magnetic domain configuration governs the stray-field landscape. Highly transparent interfaces favor stronger electronic proximity coupling, whereas reduced transparency, due to interfacial dead layers or inserted insulating spacers, can reduce electronic proximity coupling while preserving stray-field--driven effects \cite{ref11}. The magnetic domain configuration further governs these modifications, providing a handle to engineer superconducting properties via artificial or self-assembled domain patterns \cite{ref19,ref20,ref21,ref22}. However, disentangling electronic proximity effects from remanent stray-field control of vortex entry and subsequent flux flow within the same material platform remains challenging. In this context, we recently demonstrated a field-free, memory-programmable superconducting diode in structurally symmetric NbN/AlN/Py microbridges, where remanent stray fields create out-of-plane edge-field gradients that set the diode polarity and yield rectification efficiencies up to \ensuremath{\approx} 40\% \cite{ref18}. In addition, using Mo$_2$N /FePt bilayers, we achieved diode efficiencies of up to \ensuremath{\approx} 85 \% near liquid-helium temperatures \cite{ref23}.

Among the various domain structures in FM systems, stripe-like domains are particularly well suited for controlling vortex motion in superconductors. These patterns appear in materials such as FePt \cite{ref24}, Fe$_{0.2}$Ni$_{0.8}$ (permalloy, Py) \cite{ref25}, and Co/Pt multilayers \cite{ref26}, where competing anisotropies and demagnetizing fields stabilize periodic nanoscale domains, typically 50--200 nm wide. In magnetically prepared stripe states such as those considered here in FePt, the elongated domain geometry can support a spatially correlated in-plane remanent magnetization component over micrometer-scale distances. In finite geometries such as patterned electrical bridges, this extended remanence generates opposite-signed perpendicular stray-field contributions near the edges where the stripes terminate, which can influence flux entry and magnetic confinement. Importantly, stripe domains can often be rotated by an applied field and then frozen in a remanent state \cite{ref24,ref25}, enabling programmable, field-free control of the magnetic landscape during device operation \cite{ref27}. The resulting alternating out-of-plane MSF, with a well-defined periodicity, creates a spatially modulated magnetic energy landscape for vortices and can define preferential paths for vortex flow. Moreover, the coherent (dipolar) character of the remanent magnetization and its reversible rotation in hysteresis loops support the design of superconducting diodes with nonreciprocal I--V curves \cite{ref14,ref16,ref18,ref23}. Adjusting the alignment between the applied current and the stripe domains can strongly affect both the critical current density ($J_c$) and the vortex velocity in the dissipative regime \cite{ref20,ref21,ref27,ref28}. The stripe orientation changes the relation between the current-induced Lorentz force ($F_L$), the magnetic confinement near the stripe terminations, and the channels for guided vortex motion. In this way, remanent stripe-domain configurations can control the thresholds for flux injection and subsequent vortex motion, while also tuning the initial conditions for high-velocity flux flow. By controlling the entry edge and the local drive, stripe domains also shape the ensuing flux-flow regime, where a flux-flow instability can trigger a transition toward the normal state in I--V curves, as described in the Larkin-- Ovchinnikov (LO) framework \cite{ref29}. Characterizing this instability is important for determining the limits of stable flux flow, estimating the quasiparticle relaxation time (\ensuremath{\tau}) within the appropriate regime \cite{ref8,ref20}, and exploring magnon--fluxon coupling in S/FM hybrids \cite{ref30,ref31}. In practice, experimental critical velocities are often reduced by geometric imperfections and pinning centers that promote localized quenching and heating \cite{ref32,ref33,ref34,ref35,ref36}, motivating well-controlled, reconfigurable magnetic landscapes to systematically tune vortex entry and flux-flow stability.

In this work, we investigate vortex dynamics in (111)-oriented NbN thin films (10 nm) grown on Al$_2$O$_3$ substrates and capped with a nominal FePt thickness of 48 nm. We study two sample types, NbN/FePt, where FePt is deposited directly on NbN (allowing direct electronic proximity coupling as well as MSF interactions), and NbN/AlN/FePt, where a 1-2 nm thick AlN spacer reduces electronic proximity coupling while retaining magnetic coupling through stray fields \cite{ref37,ref38}. While uncapped NbN films at this thickness typically exhibit $T_c$ \ensuremath{\approx} 16 K \cite{ref39}, in our heterostructures direct contact with FePt strongly suppresses $T_c$ (\ensuremath{\approx} 9.7--9.8 K), whereas the AlN spacer restores $T_c$ close to uncapped NbN (\ensuremath{\approx} 15.5 K), consistent with proximity-driven suppression at a highly transparent interface. FePt layers thicker than about 40 nm develop nanoscale stripe domains whose in-plane orientation can be set in remanence by field-induced magnetization \cite{ref24}. This platform therefore allows us to test whether remanent magnetic control of vortex entry and transport persists when electronic proximity coupling is reduced. We probe vortex dynamics through I--V measurements while controlling the alignment between the applied current and the remanent stripe domains. This includes zero applied magnetic field, where the remanent magnetic landscape produces polarity-dependent thresholds for flux injection and vortex motion \cite{ref14,ref15}, and applied perpendicular fields, where the instability voltage and the associated apparent velocity are modified by the combined action of external and stray fields, which set the net edge field. In addition, the metallic capping layer can help redistribute localized heating associated with inhomogeneous vortex flow due to its high thermal conductivity \cite{ref35,ref40}. By comparing devices with and without the AlN spacer, we demonstrate how controllable stripe domains in NbN/FePt heterostructures enable reconfigurable nonreciprocal transport and programmable control of flux-flow stability, providing a route toward improved NbN-based cryogenic superconducting devices.

\section{Experimental Section}

NbN thin films were deposited on (0001) Al$_2$O$_3$ substrates by reactive sputtering under conditions similar to those reported in Ref. \cite{ref41}. The base pressure in the deposition chamber was 2.6\ensuremath{\times}10$^{-5}$ Pa. Substrates (0.01 \ensuremath{\times} 0.01 m$^2$) were placed 0.07 m above a Nb target (0.038 m in diameter, operated at 25 W) and maintained at 450 \textdegree{}C. The reactive atmosphere consisted of 4.25\% N$_2$ in a N$_2$+Ar mixture at a total pressure of 0.4 Pa. After deposition, the samples were cooled in the same reactive atmosphere with the heater turned off. Once at room temperature, the base pressure was re-established, and the FePt layer was grown in pure argon at 0.4 Pa using 20 W with a target distance of 0.075 m. The AlN spacer was grown by RF sputtering at a power of 25 W in an Ar/N\textsubscript{2} mixture containing 10\% N\textsubscript{2}, at a total pressure of 0.66 Pa. Substrates were mounted on the sample holder using silver paint. Deposition times were set to 2 minutes for NbN and 3 minutes for FePt, corresponding to nominal thicknesses of approximately 10 nm and 48 nm, respectively. The NbN thickness was chosen to optimize the superconducting properties of the film, maintaining a $T_c$ value comparable to optimal NbN systems ($T_c$ \ensuremath{\approx} 17 K) \cite{ref39}. Similarly, the FePt thickness was chosen to ensure the formation of a stripe domain configuration \cite{ref23,ref24,ref27}.

Structural characterization was performed by X-ray diffraction (XRD) in \ensuremath{\theta}--2\ensuremath{\theta} geometry using a Panalytical Empyrean diffractometer operated at 40 kV and 30 mA with Cu K\ensuremath{\alpha} radiation, providing an angular step of 0.013\textdegree{}. Atomic force microscopy (AFM) and magnetic force microscopy (MFM) measurements were conducted in tapping and lift modes on a Bruker Dimension 3100 microscope. Magnetization measurements were performed using a commercial magnetometer (Cryogenic Ltd.).

Electrical bridges with dimensions 40 \ensuremath{\mu}m \ensuremath{\times} 4.5 \ensuremath{\mu}m (length $L$ \ensuremath{\times} width $w$) were patterned using optical lithography \cite{ref42}. Electrical transport measurements were performed in a standard four-terminal configuration. Magnetic field--dependent measurements were carried out using two different sources: (i) a conventional electromagnet capable of generating fields up to 1 T, with in-plane and out-of-plane rotation; and (ii) an auxiliary coil producing fields up to 300 Oe, allowing for fine control of low-field conditions up to \ensuremath{\approx} 200 Oe. Current-- voltage (I-V) characteristics were measured using a Keithley nanovoltmeter (Model 2182A) and current source (Model 6221) operated in synchronized mode with 0.2 ms current pulses. No significant differences were observed when varying the pulse duration. All measurements were performed in a closed-cycle cryostat with a base temperature of 3 K, with temperature control achieved using a PID-regulated heater.

For the results presented here, the sample was first cooled to the measurement temperature. The remanent magnetic state was then prepared by applying an in-plane field of 0.5 T along the selected direction to saturate the magnetization and subsequently removing the applied field. No appreciable differences in the transport response were observed when the remanent state was instead prepared above $T_c$ before cooling. For measurements at zero applied field, the field source (coil or electromagnet) was physically moved away from the cryostat after preparation of the remanent state.

Micromagnetic simulations. Simulations were performed with MuMax3 version 3.10 \cite{ref43}, using the FePt model and remanence protocol reported for Mo$_2$N/FePt in Ref. \cite{ref23}, with the modifications introduced below. The simulated strip measured 4.096 \ensuremath{\mu}m \ensuremath{\times} 1.024 \ensuremath{\mu}m \ensuremath{\times} 50 nm and was discretized into 1024 \ensuremath{\times} 256 \ensuremath{\times} 16 cells of 4 \ensuremath{\times} 4 \ensuremath{\times} 3.125 nm$^3$. Finite boundaries along $x$ represent the bridge edges across its width, whereas periodic boundary conditions along $y$, with 20 repetitions on each side, represent the current direction when the stripes are perpendicular to the transport current. The $z$ axis is normal to the layers. A magnetic thickness of 50 nm was used to approximate the nominal 48 nm FePt layers. The calculations therefore provide a representative magnetostatic landscape rather than a quantitative reconstruction of the experimental stacks.

The material parameters were $M_s$ = $1.08 \times 10^{6}$ A/m, K$_{u1}$ = $1.9 \times 10^{5}$ J/m$^3$ with the easy axis along $z$, and $A_{ex}$ = $1.2 \times 10^{-11}$ J/m. These values give the quality factor $Q=2K_{u1}/(\mu_0 M_s^2)\approx 0.26$ and exchange length $\ell_{ex}=\sqrt{2A_{ex}/(\mu_0 M_s^2)}\approx 4.05\,\mathrm{nm}$, consistent with a weak-stripe state and cell dimensions comparable to the exchange length. The initial magnetization was nearly saturated along -x, with the normalized magnetization in cell $i$ given by $\mathbf{m}_i^{(0)}=(-\hat{\mathbf{x}}+\boldsymbol{\eta}_i)/\left|-\hat{\mathbf{x}}+\boldsymbol{\eta}_i\right|$, where the components of \ensuremath{\eta}$_i$ are independent Gaussian random variables with zero mean and standard deviation \ensuremath{\sigma} = 0.2, drawn separately for each cell. The corresponding initial dispersion in Ref. \cite{ref23} was \ensuremath{\sigma} = 0.1. The relaxation sequence followed the same procedure, with the applied field oriented along \ensuremath{-}$x$: the system was first relaxed in a 0.8 T field along \ensuremath{-}$x$ with damping \ensuremath{\alpha} = 1; the field magnitude was then reduced from 0.3 T to zero in 30 steps with \ensuremath{\alpha} = 0.5, followed by final relaxations at 1 mT along \ensuremath{-}$x$ and at zero field. These damping values were chosen for numerical relaxation and are not interpreted as physical dynamical parameters.

The perpendicular stray field was reconstructed from the remanent state using an in-house C++ code. Each source cell j was treated as a uniformly magnetized rectangular prism with magnetization $M_j = M_s\hat{m}_j$, and the field at the observation position $\mathbf r_e$ was calculated as $B_z(\mathbf r_e)=\mu_0 H_z(\mathbf r_e)=\sum_j\mathbf K_j(\mathbf r_e)\cdot\mathbf M_j$, where $K_j$ is the field kernel of cell $j$. For source-cell-center to observation-point distances $R$ \ensuremath{\leq} 50 nm, the exact rectangular-prism kernel was used; at larger distances, the point-dipole kernel was used \cite{ref23}. The evaluation plane was placed 5 nm below the lower FePt surface, within the superconducting-layer region, rather than 10 nm below the surface as in Ref. \cite{ref23}. Its distance from the centers of the nearest cell layer is therefore 5 + 3.125/2 = 6.56 nm. Superconducting screening and vortex nucleation were not included.

Longitudinal field scans follow local extrema of $B_z$ along the stripes, proceeding along x from right to left. At each position $x_k$, the transverse coordinate was obtained by Newton iteration on $\partial_y B_z=0$: $y^{(m+1)}=y^{(m)}-\frac{\partial_y B_z(x_k,y^{(m)})}{\partial_y^2 B_z(x_k,y^{(m)})}$. The iteration was seeded with the converged coordinate at the preceding scan position $x_{k-1}$. A local maximum or minimum was followed for the $\mu_0 H_z^{\mathrm{up}}$ or $\mu_0 H_z^{\mathrm{down}}$ scan, respectively. When the tracked extremum disappears, as occurs at a stripe bifurcation, the iteration converges to a neighboring extremum of the same type, producing an abrupt change in the tracked path.

\section{Results And Discussion}

\subsection{Structural, morphological, magnetic and superconducting characterization}

Structural characterization of the NbN/FePt and NbN/I/FePt samples (with I denoting the AlN spacer) was performed by XRD, as shown in Figure 1. The nominal layer thicknesses were taken to be approximately 10 nm for NbN, as previously determined by XRD and transmission electron microscopy \cite{ref38,ref41}, and 48 nm for FePt \cite{ref27}. The AlN spacer thickness was estimated to be approximately 1.6 nm \cite{ref38}. The XRD patterns show the (111) reflections of cubic NbN (PDF 03--065--9400) and cubic FePt (PDF 00--029--0717), consistent with previously demonstrated (111)-oriented NbN growth \cite{ref41}. Two main differences emerge between the samples: first, the NbN lattice parameter along the (111) direction is 0.443 nm for NbN/FePt and 0.448 nm for NbN/I/FePt, indicating a difference in the out-of-plane lattice parameter between the two heterostructures; second, the FePt (111) reflection is sharper and more intense in the sample grown directly on NbN than in the one with the AlN layer deposited at room temperature. The (111) peaks for NbN of both samples exhibit the same full width at half maximum (FWHM) of approximately 0.75\textdegree{}, which, using the Scherrer approximation (neglecting instrumental and micro-strain broadening), yields an out-of-plane coherent domain size of approximately 10 nm, consistent with the nominal NbN thickness. The samples nevertheless exhibit different out-of-plane lattice spacings and interfacial coupling \cite{ref38}.

\begin{figure}[tbp]
  \centering
  \includegraphics[width=\columnwidth,height=0.58\textheight,keepaspectratio]{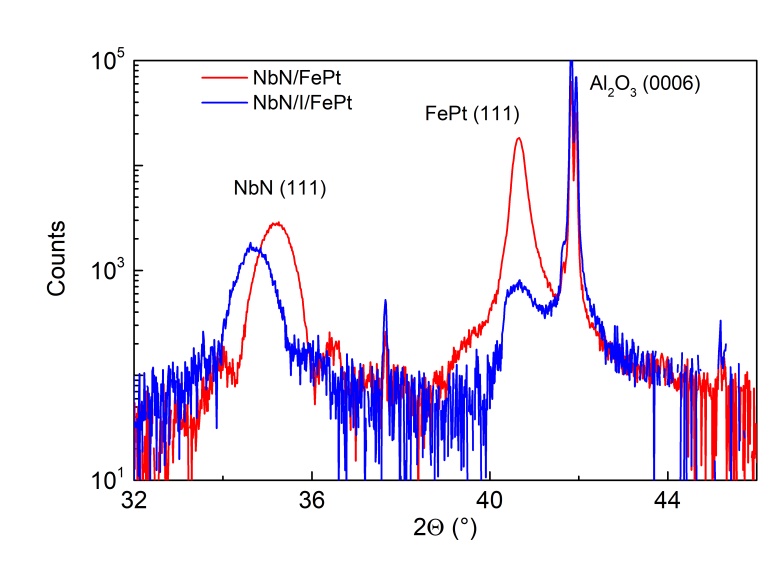}
  \caption{X-ray diffraction (XRD) patterns of NbN/FePt and NbN/I/FePt heterostructures showing NbN(111) and FePt(111) reflections.}
  \label{fig:01}
\end{figure}

To characterize the sample surface both morphologically and magnetically, AFM and MFM measurements were performed at room temperature. Figures 2a and 2b show representative 1 \ensuremath{\times} 1 \ensuremath{\mu}m$^2$ images of NbN/I/FePt. The displayed scale emphasizes the nanoscale stripe domains in the FePt layer. The surface is very smooth, with occasional three-dimensional defects that appear isolated in larger-area scans (height of about 3 nm, as seen in the height profile along the dashed line in Figure 2a). The root mean square (RMS) roughness is approximately 0.3 nm, consistent with a flat NbN surface and homogeneous FePt growth at room temperature \cite{ref27,ref41}. In addition, the MFM image reveals the expected stripe-domain structure characteristic of FePt layers thicker than 40 nm \cite{ref24}. The stripe half-period (stripe width) is about 50 nm, while the full magnetic periodicity is approximately 100 nm, as extracted from the phase profile in Figure 2b. This periodicity, used as a characteristic length scale, when compared with the intervortex distance in an Abrikosov lattice of $a=\sqrt{2\Phi_0/(\sqrt{3}\mu_0 H)}$, corresponds to an equivalent field scale of $\approx$ 0.24 T, indicating that the domain periodicity is comparable to the intervortex spacing within the field range explored below.

\begin{figure}[tbp]
  \centering
  \includegraphics[width=\columnwidth,height=0.58\textheight,keepaspectratio]{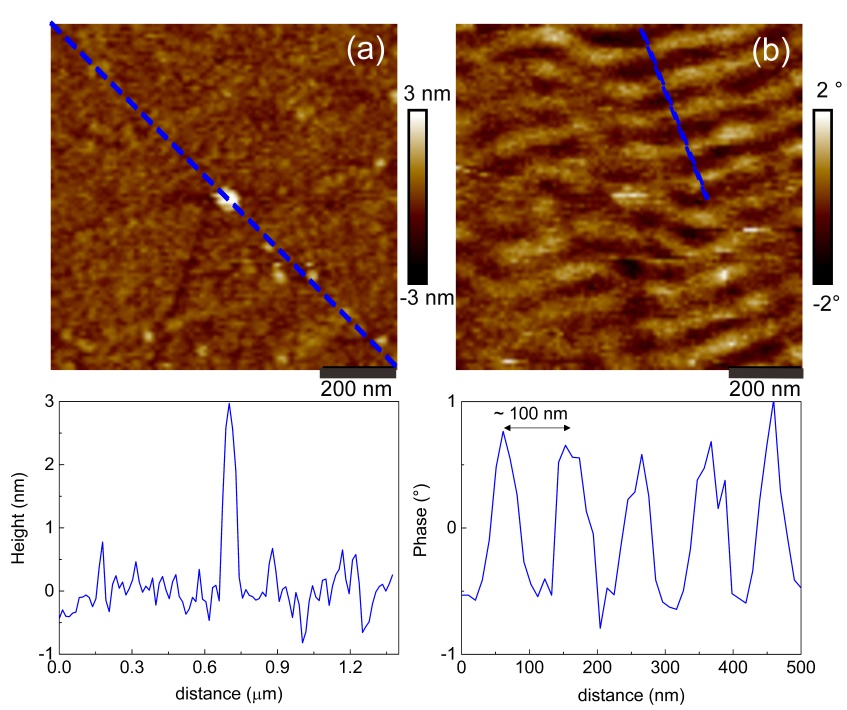}
  \caption{Morphological and magnetic characterization of the NbN/I/FePt surface. (a) AFM height image (1 \ensuremath{\times} 1 \ensuremath{\mu}m$^2$) with RMS roughness of 0.3 nm. (b) MFM phase image (1 \ensuremath{\times} 1 \ensuremath{\mu}m$^2$) highlighting nanoscale stripe domains with ~50 nm lateral width and ~100 nm peak-to-peak periodicity.}
  \label{fig:02}
\end{figure}

Micromagnetic simulations quantify the remanent FePt stray-field landscape and its variations near the bridge edges and internal stripe defects (Fig. 3). The model and numerical protocol are based on those used for Mo$_2$N/FePt \cite{ref23}, with the magnetic thickness reduced from 56 to 50 nm, the initial Gaussian dispersion increased from \ensuremath{\sigma} = 0.1 to 0.2, and the field evaluated 5 nm rather than 10 nm below the FePt surface. The present remanence is directed along \ensuremath{-}$x$. The resulting calculation allows the edge-field asymmetry and the effect of a stripe bifurcation to be examined within the same model. The full numerical procedure is given in the Experimental Section.

Figure 3a shows the relaxed remanent configuration at the mid-thickness of the FePt layer. Dark and light stripes correspond to positive (+$z$) and negative (\ensuremath{-}$z$) perpendicular magnetization components, respectively. The stripe axes lie along $x$, parallel to the remanent magnetization directed along \ensuremath{-}$x$, and the texture is modulated along y. At the transverse cut shown in Fig. 3b, nine periods fit within the 1.024 \ensuremath{\mu}m periodic cell, giving 18 alternating domains and a period $\Delta_s$ = 1024/9 \ensuremath{\approx} 114 nm. Periodic boundary conditions restrict the allowed periods to (1024 nm)/$n$, with integer $n$. The selected value is the closest allowed period to the approximately 110 nm MFM value reported for Mo$_2$N/FePt \cite{ref23}. The relaxed pattern also contains two stripe bifurcations of opposite character. Near $x$ \ensuremath{\approx} 1200 nm and y \ensuremath{\approx} 650 nm a stripe splits, adding one period to the cell, whereas near $x$ \ensuremath{\approx} 2900 nm, at the periodic boundary ($y$ \ensuremath{\approx} 0 and 1024 nm), two stripes merge and remove it. Between the two defects the cell therefore accommodates ten periods (\ensuremath{\approx} 102 nm), close to the approximately 100 nm periodicity of the present samples (Fig. 2b). Such defects are consistent with the stripe forks and terminations visible in the MFM texture.

Figures 3b--d show the perpendicular stray field $B_z$ = $\mu_0 H_z$ in the superconducting-layer region, 5 nm below the FePt surface. The transverse cut in Fig. 3b, taken in a regular region at $x$ \ensuremath{\approx} 500 nm, is nearly sinusoidal in $y$, with the stripe period and extrema of approximately \ensuremath{-}200 and +185 mT. Outside the magnetic surface, a Fourier component with wave number $k_n$ = 2\ensuremath{\pi}n/$\Delta_s$ decays as $B_n(d)=B_n(0)e^{-k_n d}$, where d denotes distance from the magnetic surface in this expression. At $d=5$ nm, propagation attenuates the third harmonic relative to the fundamental by $e^{-4\pi d/\Delta_s}\approx 0.58$. This filtering is moderate at the evaluation distance, consistent with the nearly sinusoidal profile reflecting both the weak-stripe magnetization texture and attenuation of higher spatial harmonics.

The horizontal scans in Figs. 3c and 3d follow local maxima ($\mu_0 H_z^{\mathrm{up}}$, red) and minima ($\mu_0 H_z^{\mathrm{down}}$, blue) of the field, using the procedure described in the Experimental Section. Figure 3c corresponds to the dotted paths in the lower part of Fig. 3a, where the stripes are approximately straight and free of major defects. The positive scan remains positive and the negative scan remains negative throughout the strip. Both profiles rise from their left-edge values toward broad interior plateaus of approximately +180 and \ensuremath{-}180 mT, respectively, and rise again on approaching the right edge. Thus, the field magnitude reaches approximately 365 mT near the left edge for the negative scan and near the right edge for the positive scan, while it is reduced at the opposite edge. More precisely, the plateau magnitudes are approximately 185 mT where nine periods fit in the periodic cell ($x$ \ensuremath{\lesssim} 1200 nm and x \ensuremath{\gtrsim} 2900 nm) and approximately 175 mT between the two bifurcations, where the local period is \ensuremath{\approx} 102 nm. At $x$ \ensuremath{\approx} 1200 and 2900 nm both scans change in magnitude by similar amounts, as expected from the change in local period rather than from the edge correction, which shifts both scans in the same direction.

\begin{figure}[tbp]
  \centering
  \includegraphics[width=\columnwidth,height=0.58\textheight,keepaspectratio]{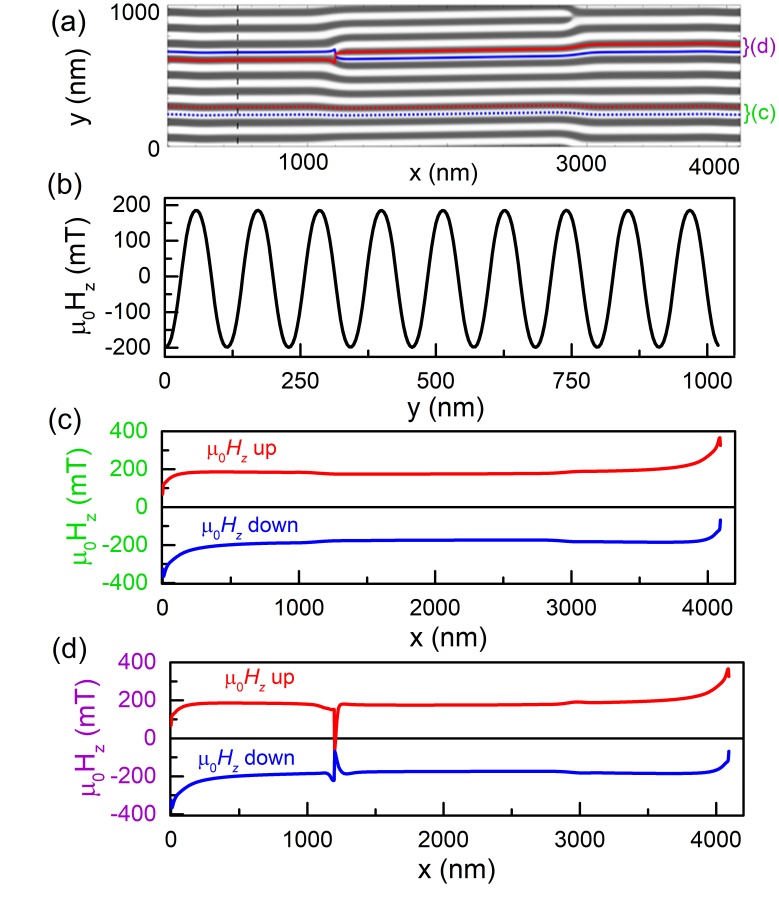}
  \caption{Micromagnetic configuration and calculated FePt stray field for remanence along \ensuremath{-} $x$, obtained by relaxing a nearly saturated state with Gaussian dispersion \ensuremath{\sigma} = 0.2. (a) Normalized perpendicular magnetization $m_z$ at the mid-thickness of the 4096 \ensuremath{\times} 1024 \ensuremath{\times} 50 nm$^3$ simulated strip. Dark and light stripes correspond to positive and negative perpendicular magnetization components, respectively. Red and blue paths follow local field maxima and minima. Dotted paths lie in a regular region [profiles in (c)], whereas solid paths cross the bifurcation near $x$ \ensuremath{\approx} 1200 nm [profiles in (d)]. The dashed line marks the transverse cut at $x$ \ensuremath{\approx} 500 nm. (b) Transverse field profile with $\Delta_s$ \ensuremath{\approx} 114 nm and extrema of approximately \ensuremath{-}200 and +185 mT. (c) $\mu_0 H_z^{\mathrm{up}}$ and $\mu_0 H_z^{\mathrm{down}}$ along the dotted paths, showing interior plateaus and asymmetric edge enhancement. (d) Corresponding profiles along the solid paths, showing the local variations and path switches at the bifurcation. All fields are evaluated 5 nm below the lower FePt surface, in the absence of superconducting screening, and are reported in mT.}
  \label{fig:03}
\end{figure}

The edge corrections are consistent with the remanent magnetization along \ensuremath{-}$x$. The in-plane component produces surface magnetic charges $\sigma_m=\mathbf M\cdot\hat{\mathbf n}$ on the faces normal to $x$. For remanence along \ensuremath{-}$x$, the right face carries negative charge and the left face positive charge. In the observation plane below FePt, their field adds a correction of the same sign to both stripe-polarity scans: negative near the left edge and positive near the right edge. Reversal of this edge correction provides a magnetostatic basis for reversal of the diode polarity.

Figure 3d isolates the influence of the bifurcation through the solid paths in the upper part of Fig. 3a. Away from the defect, the profiles resemble those in Fig. 3c. Near $x$ \ensuremath{\approx} 1200 nm, however, $\mu_0 H_z^{\mathrm{up}}$ drops sharply and becomes negative over a narrow interval, reaching approximately \ensuremath{-}60 mT, whereas $\mu_0 H_z^{\mathrm{down}}$ first deepens to approximately \ensuremath{-}220 mT and then develops a narrow upward excursion to approximately \ensuremath{-}70 mT while remaining negative. At the bifurcation, each tracked extremum disappears and the corresponding path switches to a neighboring extremum of the same type, as shown in Fig. 3a. These path switches contribute to the abrupt features of the scans and should be distinguished from spatial discontinuities of the physical field. The local field variations identify an interior region where magnetic gradients can provide confinement and pinning for vortices or antivortices \cite{ref44}, but the scans alone do not quantify the pinning force or depinning current. A finite density of such stripe defects can therefore add interior pinning sites to the edge-controlled magnetic landscape.

Compared with Ref. \cite{ref23}, the alignment of stripes with the remanence, the nearly sinusoidal transverse modulation, the opposite-signed interior plateaus, and the remanence-dependent edge correction are preserved. In Ref. \cite{ref23}, the relaxed state contained eight periods within the 1.024 \ensuremath{\mu}m periodic cell ($n$ = 8, $\Delta_s$ = 128 nm). The selected period and stripe-defect pattern differ between the two calculations, in which both the magnetic thickness and the initial dispersion were changed. The field magnitudes are approximately 180 mT in the interior and 365 mT near the enhanced edges, compared with approximately 160 and 280 mT in Ref. \cite{ref23}. Evaluating the field closer to FePt reduces the spatial attenuation, but the effects of the changed thickness and magnetic texture are not isolated by this comparison. The persistence of the main field features across these two calculations supports the qualitative robustness of the edge-controlled magnetostatic picture. The calculation describes the field generated by FePt. Because it omits superconducting screening and vortex nucleation, it does not establish the actual vortex population. The possible contributions of internal vortex confinement and activation to the measured dissipative branches are considered below.

We now turn to the low-temperature transport and magnetic measurements. Figure 4a shows low-temperature electrical transport measurements. The inset displays an image of a typical microbridge used in this study. The data reveal that $T_c$ in NbN/I/FePt is approximately 15.5 K, close to that of uncapped NbN thin films \cite{ref41}, while in NbN/FePt it is reduced to 9.7-9.8 K. This suppression has previously been attributed to the high transparency of the SC/FM interface, which promotes strong electronic proximity coupling and suppresses superconductivity over a length scale comparable to the coherence length \ensuremath{\xi}. Comparison with single films suggests an effective superconducting thickness below 5 nm \cite{ref38,ref45}. Figure 4b displays magnetic hysteresis loops at 20 K with the field applied parallel to the substrate (macroscopic samples). These measurements were performed above $T_c$ to exclude superconducting contributions. The data show the expected FM signal with a saturation magnetization near 1100 emu/cm$^3$, along with characteristic features of stripe-domain structures, such as an approximately linear magnetization response after the hysteresis region and before saturation. Both samples exhibit a sharp magnetization reversal at the coercive field, consistent with a largely coherent in-plane reorientation of the magnetization. The remanence ($M_r$) and the coercive field ($H_c$) depend on the sample, with values of approximately 0.65 \ensuremath{\times} $M_s$ and 0.5 \ensuremath{\times} $M_s$, and $\mu_0 H_c$ of \ensuremath{\approx} 26 mT and 18 mT for NbN/FePt and NbN/I/FePt, respectively. Both stacks display a similar saturation magnetization, indicating that the saturation magnetic moment per unit FePt volume is comparable. However, the reduced remanence in NbN/I/FePt suggests a reduced coherent (dipolar) remanent component, which may weaken the remanent edge stray fields relevant for vortex entry. In the following, these differences must therefore be considered together with the changes in interfacial coupling introduced by the AlN spacer. These differences may be related to changes in microstructure and the fact that FePt is more strongly textured along the (111) axis when grown directly on the epitaxial NbN layer.

\begin{figure}[tbp]
  \centering
  \includegraphics[width=\columnwidth,height=0.58\textheight,keepaspectratio]{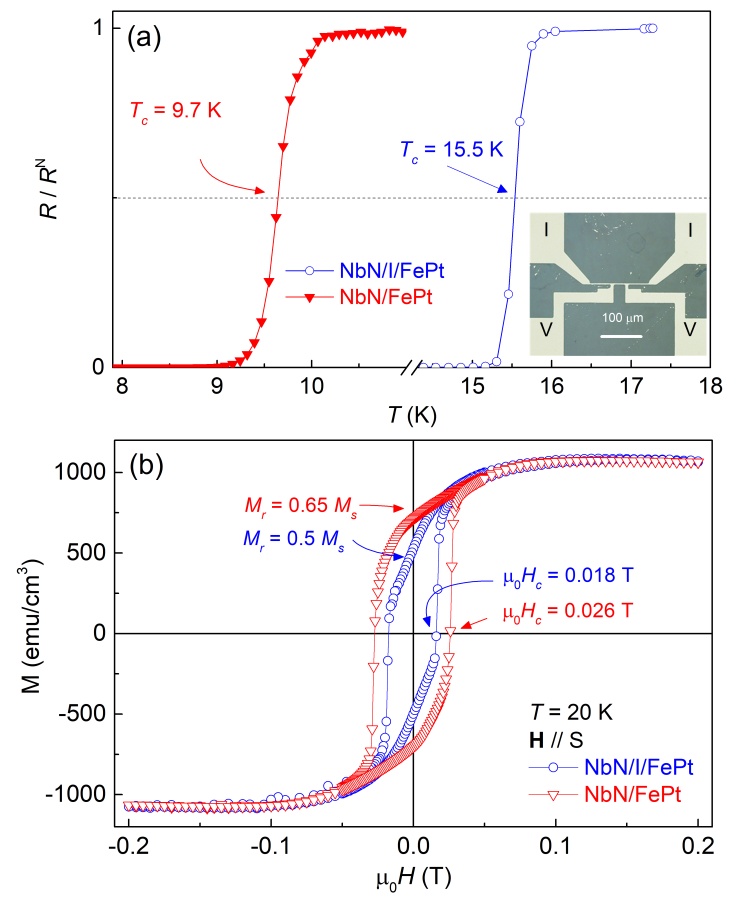}
  \caption{Low-temperature transport and magnetic characterization. (a) Normalized resistance vs. temperature for NbN/FePt and NbN/I/FePt. The inset shows a typical microbridge geometry indicating voltage (V) and current (I) terminals. (b) Magnetic hysteresis loops at 20 K with the field applied parallel to the substrate plane.}
  \label{fig:04}
\end{figure}

\subsection{Magnetic landscape configuration and nonreciprocal transport}

Figure 5 combines MFM observations of patterned microbridges with schematic representations of the remanent magnetic landscape and its proposed influence on vortex dynamics. Figures 5a--c show MFM phase images obtained after applying and removing a saturating in-plane magnetic field along different directions \cite{ref24,ref25,ref27}. The darker central regions correspond to the microbridges, where well-defined stripe domains are preserved after patterning. By changing the direction of the preparation field, the stripes can be oriented at approximately 90\textdegree{}, 45\textdegree{}, and 0\textdegree{} relative to the current direction, respectively. Here, the stripe--current angle \ensuremath{\theta} is defined as the smaller angle between the stripe axis and the bridge longitudinal axis, independently of the current polarity. It is important to note that \ensuremath{\theta} = 0\textdegree{} and 180\textdegree{} describe geometrically equivalent stripe orientations and pinning landscapes. Figure 5d illustrates configurations I and II, defined by opposite remanent magnetizations at the same stripe orientation, shown here for \ensuremath{\theta} = 90\textdegree{}, while maintaining the same current sign convention. Importantly, the simulated remanence in Fig. 3 corresponds to configuration I, with a negative edge-field correction on the left and a positive correction on the right. The reproducible rotation and reversal of the magnetic texture are consistent with the sharp magnetization reversal observed in the hysteresis loops (Fig. 4b). The physical origin of the nonreciprocal response is illustrated schematically in Figures 5e and 5f. In a simplified dipolar approximation, neglecting additional local out-of-plane stray-field contributions arising from irregularities in the stripe pattern, the remanent in-plane magnetization generates opposite-signed perpendicular stray-field components near the two bridge edges. For configuration I (Fig. 5e), the macroscopic edge contributions are negative at the left edge and positive at the right edge. These contributions are superimposed on the local periodic stray-field modulation associated with the stripe domains. Within the proposed edge-entry picture \cite{ref14,ref23}, this magnetic landscape favours the nucleation and confinement of antivortices and vortices near the left and right edges, respectively. As illustrated in Figure 5f, negative current drives these preferential edge populations toward the bridge interior through the $F_L$. Once the magnetic confinement is overcome, guided vortex motion along the stripe channels, potentially accompanied by vortex--antivortex annihilation, can produce dissipation at relatively low current. Conversely, positive current drives these flux polarities outward, opposing inward flux injection and favoring a vortex-depleted central region associated with the higher-critical-current branch. Reversing the remanent magnetization exchanges the edge-field polarities and consequently reverses the diode polarity.

The stripe orientation provides an additional degree of control over vortex transport. When the stripes are perpendicular to the current (\ensuremath{\theta} = 90\textdegree{}), the $F_L$ is directed along the stripe channels, favouring guided vortex motion. In contrast, for \ensuremath{\theta} = 0\textdegree{}, vortex propagation requires crossing the periodic magnetic modulation, while intermediate orientations involve both contributions \cite{ref21,ref27}. Thus, rotating the stripe domains modifies both the edge-injection conditions and the available paths for vortex motion. Importantly, the vortex distributions depicted in Figure 5f represent an idealized physical scenario rather than directly measured configurations. In particular, the transport measurements do not establish a completely vortex-free interior. Nevertheless, the proposed mechanism provides a consistent framework for understanding the polarity-dependent dissipation and the angular control of the superconducting diode response discussed below.

\begin{figure*}[tp]
  \centering
  \includegraphics[width=0.91\textwidth,height=0.56\textheight,keepaspectratio]{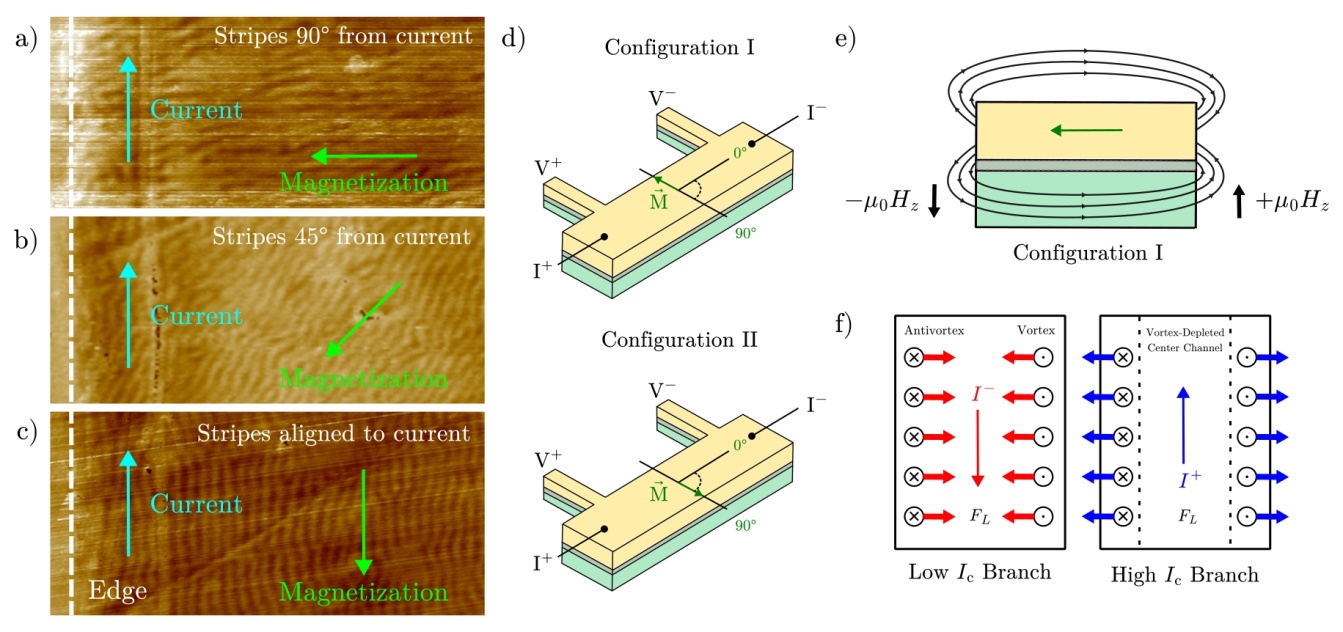}
  \caption{Remanent stripe-domain configurations and proposed vortex dynamics in NbN/FePt microbridges. (a--c) MFM phase images after in-plane magnetic saturation and field removal, showing stripe orientations of 90\textdegree{}, 45\textdegree{}, and 0\textdegree{} (equivalent to 180\textdegree{}) relative to the current direction, respectively. (d) Configurations I and II, with opposite remanent in-plane magnetizations and stripes perpendicular to the current. (e) Dipolar approximation of the MSF in configuration I, illustrating opposite out-of-plane field components near the bridge edges. (f) Proposed polarity-dependent vortex dynamics in configuration I following an ideal dipolar-like behaviour: negative current drives edge vortices and antivortices inward, promoting guided flux motion and possible annihilation (low-$I_c$ branch), whereas positive current drives them outward, suppressing flux injection and favoring a vortex-depleted central region (high-$I_c$ branch). No internal vortices are considered in this picture of the edge-mediated mechanism.}
  \label{fig:05}
\end{figure*}

To investigate the superconducting diode effect in these heterostructures, we consider the competition between the current-induced $F_L$ and magnetic confinement associated with the remanent MSF of the FM capping layer \cite{ref14,ref46,ref47}. The transport self-field (SF) provides a secondary correction to this magnetic landscape. To estimate its scale at the sample edges, we use the thin-strip expression for an approximately uniform current density \cite{ref47}: $B_{\mathrm{edge}}(I)\approx \frac{\mu_0 I}{2\pi w}\left[\ln\left(\frac{2w}{d}\right)+1\right]$, where $I$ is the applied current magnitude and w is the bridge width. For a strip with w = 4.5 \ensuremath{\mu}m and d = 10 nm, the edge SF scales as \ensuremath{\approx} 0.35 mT mA$^{-1}$, i.e., \ensuremath{\approx} 1.7 mT at 5 mA. This value is much smaller than the edge stray fields calculated in Fig. 3, obtained from micromagnetic simulations assuming a laterally coherent in-plane remanent magnetization component across the bridge width, for which the maximum field magnitude reaches \ensuremath{\approx} 365 mT.

Following the edge-injection framework proposed by Carapella $et$ $al.$ \cite{ref14} and subsequently applied to Mo$_2$N/FePt \cite{ref23}, we interpret the transport response in terms of preferential vortex and antivortex populations near opposite bridge edges, favored by the remanent stray-field contributions (Fig. 5). This preference is not exclusive, since the local stripe modulation allows both signs of the perpendicular stray field at either edge. For a regular stripe configuration, we assume that spontaneous vortex--antivortex nucleation occurs predominantly near the bridge edges, leaving the regular interior initially free of vortices. One current polarity drives the preferential edge populations inward, enabling guided motion and vortex--antivortex annihilation once magnetic confinement is overcome, and producing dissipation at relatively low current. The opposite polarity drives these populations outward and opposes inward injection, maintaining a vortex-free central channel in the ideal regime. Reversing the remanent magnetization exchanges these roles and reverses the diode polarity. Rotating the stripe axis relative to the current changes both the remanent edge-field contribution and the alignment between the magnetic channels and the $F_L$, consistent with the suppression of nonreciprocity toward the parallel-stripe configuration. We denote the stripe--current angle by \ensuremath{\theta}, as defined in Fig. 5, and use the same notation throughout Figs. 6--9. The vortex-free interior remains an assumption of this idealized picture: transport alone cannot distinguish such a channel from one containing immobilized vortices. However, as discussed below, the low dissipation threshold for the polarity that drives edge vortices and antivortices inward contrasts with the much higher threshold of the opposite branch. Despite the sizable transverse modulation of the perpendicular stray field, this contrast suggests that any pre-existing interior vortices contribute only weakly to dissipation in the high-critical-current branch. For convenience, we quantify the nonreciprocal regime through $\Delta I_c=|I_c^+-I_c^-|$, with $I_c^\pm=|I_c(\pm)|>0$. In addition, we define the diode efficiency as $\eta=\frac{|I_c^+-I_c^-|}{I_c^++I_c^-}$, consistent with the efficiencies quoted in the Introduction. We next examine the temperature and angular dependence of the diode response using remanent stripe configurations prepared by applying and removing an in-plane magnetic field at selected angles.

Figure 6 summarizes the transport results for NbN/FePt. Panel (a) shows I--V curves at 4.5 K for different angles (\ensuremath{\theta}) between the current and the stripe-domain axis, with configuration I in Fig. 5 serving as the reference remanent state. Panel (b) presents the corresponding (|V|) versus (|I|) curves on log--log scales for selected angles, and panel (c) shows the diode efficiency \ensuremath{\eta} as a function of \ensuremath{\theta}. The curves exhibit pronounced nonreciprocity that depends on the remanent stripe orientation. Both current polarities develop finite voltage before switching, although their dissipation thresholds differ substantially. The critical current Ic is determined using a voltage criterion of 5 \ensuremath{\mu}V, slightly above the noise level. For \ensuremath{\theta} = 90\textdegree{}, the larger critical current reaches approximately 1.5 mA, with a diode efficiency of about 40--45\%. The angular dependence shows a broad region of elevated efficiency at large stripe--current angles and strong suppression as the stripes approach alignment with the current. Panels (d) and (e) show the temperature-dependent I-- V characteristics for \ensuremath{\theta} = 90\textdegree{}, while panel (f) compares \ensuremath{\eta} (T) for \ensuremath{\theta} = 90\textdegree{}, \ensuremath{\theta} = 45\textdegree{}, and \ensuremath{\theta} = 0. For \ensuremath{\theta} = 90\textdegree{}, the efficiency increases overall with temperature, reaching approximately 60\% at 8--9 K, near $T_c$. A similar but weaker increase occurs for \ensuremath{\theta} = 45\textdegree{}, where the efficiency approaches 50\% at the highest measured temperatures. For \ensuremath{\theta} = 0\textdegree{}, the efficiency remains lower but exhibits appreciable temperature dependence, reaching approximately 20\% around 8 K. These trends are consistent with the proposed role of remanent edge fields in controlling flux injection and of stripe channels in guiding vortex motion. As $T$ approaches $T_c$, reduced vortex pinning may facilitate motion along these channels and contribute to the observed increase in \ensuremath{\eta}, as discussed previously for Mo$_2$N/FePt \cite{ref23}. We next compare these results with NbN/AlN/FePt to examine how the AlN spacer affects the angular and temperature dependence of the diode response.

\begin{figure*}[tp]
  \centering
  \includegraphics[width=0.91\textwidth,height=0.56\textheight,keepaspectratio]{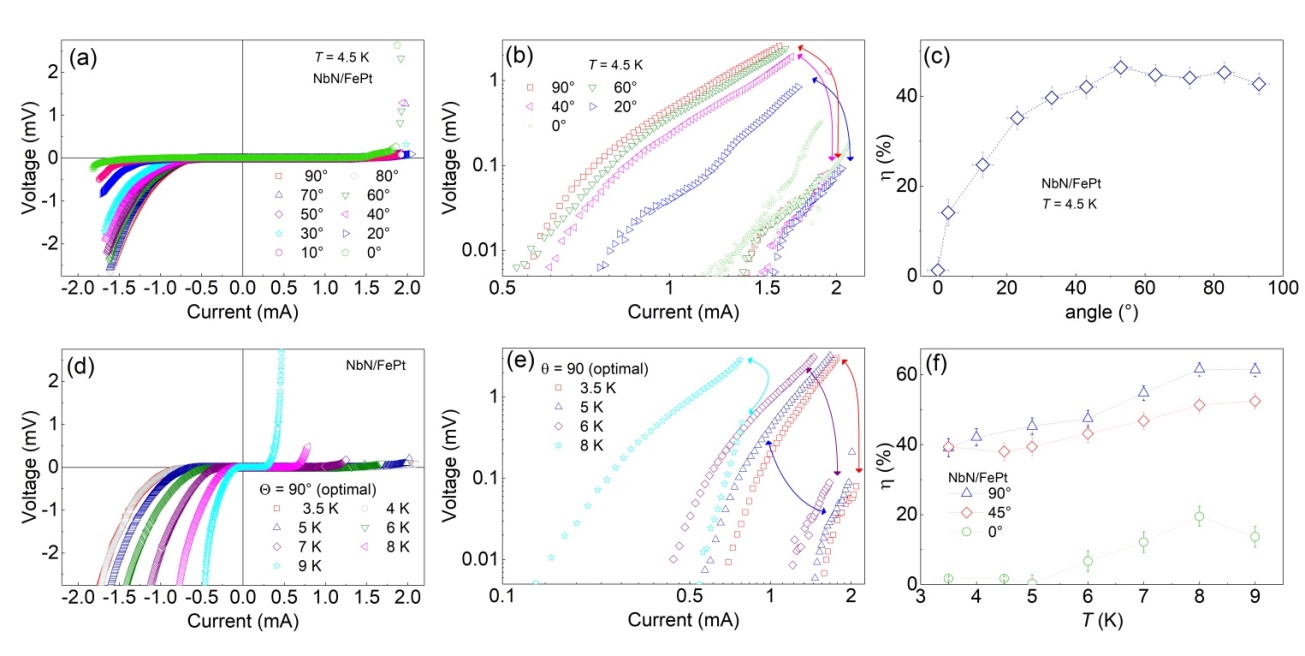}
  \caption{Angle- and temperature-dependent nonreciprocal vortex dynamics in NbN/FePt. (a) Current-- voltage (I--V) characteristic curves at 4.5 K for various orientations \ensuremath{\theta} between the transport current and the remanent stripe-domain direction, illustrating the angular control of nonreciprocal dissipation following configuration I of Fig. 5. (b) Log--log plot of the positive and negative voltage branches (|V| vs |I|) for selected angles, emphasizing the asymmetric onset of dissipation. The critical current $I_c$ is determined using a voltage criterion of 5 \ensuremath{\mu}V, slightly above the noise level. (c) Diode efficiency as a function of \ensuremath{\theta} at 4.5 K, showing a small nonreciprocity for \ensuremath{\theta} \ensuremath{\approx} 0\textdegree{} and a maximum of about 40--50\% when the stripes are nearly perpendicular to the current (\ensuremath{\theta} \ensuremath{\approx} 60--90\textdegree{}). Error bars were estimated from the uncertainty in $I_c$ associated with extrapolating a linear fit of log |V| versus log |I| to the 5 \ensuremath{\mu}V voltage criterion. (d) I--V curves measured at various temperatures for the optimal stripe configuration (\ensuremath{\theta} = 90\textdegree{}), showing a decrease in the characteristic current scales with increasing T. (e) Log--log plots of |V| vs |I| for the curves in (d), highlighting the temperature dependence of the asymmetry. (f) Diode efficiency as a function of temperature for different domain orientations (\ensuremath{\theta} = 0\textdegree{}, 45\textdegree{}, and 90\textdegree{}), confirming that the nonreciprocal response is robust and largest when the stripes are perpendicular to the current.}
  \label{fig:06}
\end{figure*}

Figure 7 summarizes the corresponding measurements for NbN/AlN/FePt. Panels (a), (b), and (c) show I--V curves at 4.5 K for different angles \ensuremath{\theta} between the current and the remanent stripe-domain axis, the associated |V| versus |I| plots on log--log scales for selected angles, and the diode efficiency \ensuremath{\eta}(\ensuremath{\theta}) following configuration I of Fig. 5, respectively. As in NbN/FePt, the nonreciprocal response is weak when the stripes are aligned with the current and becomes more pronounced as they are rotated towards \ensuremath{\theta} = 90\textdegree{}. In this configuration, the larger critical current reaches approximately 4 mA and \ensuremath{\eta} attains about 20--25\%. The persistence of the diode response with the AlN spacer supports the role of remanent stray-field coupling in a structure with reduced electronic proximity effects and higher $T_c$. The lower remanent magnetization of NbN/AlN/FePt (Fig. 4b) may contribute to its smaller efficiency by reducing the remanent edge-field amplitude. However, comparisons at the same absolute temperature involve different reduced temperatures $T$ / $T_c$ and potentially different vortex pinning strengths, so the efficiency difference cannot be attributed to the remanent magnetization alone \cite{ref48}. Panels (d) and (e) present the temperature-dependent I--V characteristics for \ensuremath{\theta} = 90\textdegree{}, while panel (f) compares \ensuremath{\eta}($T$) for \ensuremath{\theta} = 0\textdegree{} and \ensuremath{\theta} = 90\textdegree{}. The current asymmetry persists over a broad temperature range. For \ensuremath{\theta} = 90\textdegree{}, the efficiency increases overall with temperature, reaching approximately 35--40\% at the highest measured temperatures, with some nonmonotonic variation. For \ensuremath{\theta} = 0\textdegree{}, it remains below approximately 10\% throughout the measured range. The increase in \ensuremath{\eta} reflects a growing relative asymmetry between the critical-current magnitudes; it does not by itself imply an increase in the absolute current window. Within the proposed picture, stronger vortex pinning could raise the current required for motion along the stripe channels and contribute to the smaller efficiency of NbN/AlN/FePt. Thermal weakening of pinning as $T$ approaches $T_c$ may then facilitate guided vortex motion and enhance the contrast between the two current polarities.

The comparison of Figs. 6 and 7 with our previous results on Mo$_2$N/FePt \cite{ref23} reveals a common polarity-dependent transport mechanism, but with important differences in the onset of dissipation. In Mo$_2$N/FePt, one current polarity remains nearly dissipationless up to abrupt switching, whereas the opposite polarity develops finite voltage at much lower currents. In contrast, both branches of the NbN-based devices exhibit dissipation before switching, and even the lower-critical-current branch retains a substantial current threshold. This difference suggests a more pronounced contribution from vortex pinning in NbN, consistent with previous measurements on related epitaxial films \cite{ref48}. Thus, although the stripe domains provide preferential channels for vortex motion \cite{ref27}, propagation along these channels remains influenced by the intrinsic pinning landscape of the superconducting layer.

\begin{figure*}[tp]
  \centering
  \includegraphics[width=0.91\textwidth,height=0.56\textheight,keepaspectratio]{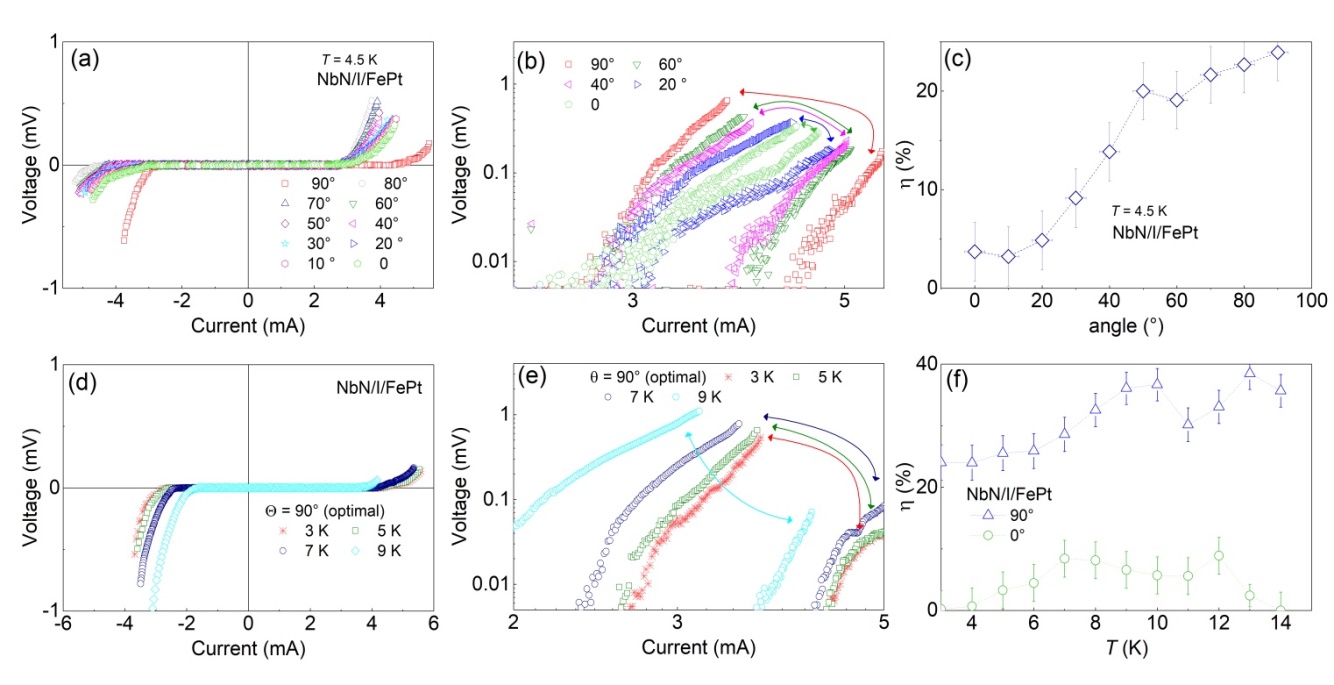}
  \caption{Angular and thermal tuning of nonreciprocal vortex dynamics in NbN/I/FePt. (a) I--V curves at 4.5 K for different domain orientations \ensuremath{\theta} following configuration I of Fig. 5, showing strong rectification for \ensuremath{\theta} = 90\textdegree{}, with the stripes perpendicular to the current. (b) Log--log plot of the positive and negative voltage branches (|V| vs |I|) for selected angles, highlighting the asymmetry in vortex dissipation. (c) Diode efficiency \ensuremath{\eta} as a function of \ensuremath{\theta} at 4.5 K, with a maximum near \ensuremath{\theta} = 90\textdegree{}, consistent with the proposed remanent stray-field origin of the effect. (d) I--V curves for increasing temperature at the optimal configuration (\ensuremath{\theta} = 90\textdegree{}), showing reduced characteristic current scales as temperature increases. (e) Log--log plots from (d), emphasizing the temperature dependence of the dissipation asymmetry. (f) Temperature dependence of the diode efficiency for \ensuremath{\theta} = 0\textdegree{} and \ensuremath{\theta} = 90\textdegree{}, showing a pronounced nonreciprocal response over a wide temperature range when the domain orientation maximizes the edge MSF, and a smaller \ensuremath{\eta} for \ensuremath{\theta} = 0\textdegree{}.}
  \label{fig:07}
\end{figure*}

The pronounced asymmetry between current polarities also constrains the possible vortex distribution within the microbridge. For stripes perpendicular to the current (\ensuremath{\theta} = 90\textdegree{}), the $F_L$ acts along the stripe channels. A substantial population of weakly pinned vortices and antivortices within the regular stripe regions would be expected to contribute to dissipation for both current polarities, reducing the contrast between their dissipation thresholds. The observed nonreciprocity therefore suggests that vortex transport is not dominated by a pre-existing population of readily mobile vortices in the regular stripe regions. However, this does not exclude vortices localized at internal magnetic defects. Stripe bifurcations, as illustrated in Fig. 3, and internal stripe terminations can generate local stray-field inhomogeneities that may favour vortex nucleation and confinement. Within the proposed mechanism, negative current in configuration I favours inward injection of vortices and antivortices from opposite edges, followed by motion along the stripe channels once the relevant entry and pinning barriers are overcome. Positive current instead opposes this injection, while vortices trapped at internal magnetic defects may remain immobilized up to higher currents. Their subsequent activation, or residual vortex injection from the edges, could account for the finite-voltage response of the higher-critical-current branch. The transport measurements cannot distinguish these contributions or establish whether the interior is entirely vortex-free.

To further assess the physical plausibility of this scenario, we estimate the characteristic superconducting length and field scales using NbN/AlN/FePt as a reference, since the AlN spacer reduces electronic proximity effects and preserves $T_c$ \ensuremath{\approx} 15.5 K. The zero-temperature penetration depth is estimated using the dirty-limit relation $\lambda(0)=\sqrt{\hbar\rho_N/(\pi\mu_0\Delta_0)}$. Taking $\rho_N$ \ensuremath{\approx} 80 \ensuremath{\mu}\ensuremath{\Omega}\ensuremath{\cdot}cm \cite{ref41} and $\Delta_0=2.1k_B T_c$ \cite{ref49}, with $T_c$ = 15.5 K, gives \ensuremath{\lambda}(0) \ensuremath{\approx} 220 nm. Using a representative coherence length of \ensuremath{\xi} (0) \ensuremath{\approx} 5.5 nm reported for comparable NbN films \cite{ref41}, we estimate $\mu_0 H_{c1}$ \ensuremath{\approx} 15 mT. In the Py/Nb system studied by Carapella $et$ $al.$ \cite{ref14}, the estimated lower critical field exceeded the interior stray-field amplitude, supporting preferential vortex nucleation near the bridge edges. In the micromagnetic calculation (Fig. 3), the edge stray fields reach approximately 365 mT, considerably exceeding $\mu_0 H_{c1}$. However, this comparison does not establish whether spontaneous vortices also populate the regular stripe regions. The corresponding Pearl length, \ensuremath{\Lambda} = 2$\lambda^2$(0)/$d$ > 9 \ensuremath{\mu}m, exceeds the bridge width of 4.5 \ensuremath{\mu}m, while the approximately 50 nm stripe width is substantially smaller than \ensuremath{\lambda}(0). Under these conditions, vortex nucleation cannot be inferred from a simple comparison between the local magnetic field and the bulk lower critical field. Indeed, Bobba $et$ $al.$ \cite{ref50} demonstrated that spontaneous vortex--antivortex formation in S/FM hybrids depends strongly on the superconducting thickness, magnetic domain width, and out-of-plane magnetization. Furthermore, Di Giorgio $et$ $al.$ \cite{ref51} experimentally observed localized vortex nucleation at magnetic bifurcations even when the regular stripe regions remained free of spontaneous vortices. These findings support the possibility that the regular stripe regions in our devices contain few or no spontaneous vortices, while localized vortices may be stabilized at magnetic defects. Such a distribution would be compatible with the strong polarity dependence of the observed dissipation, although its existence cannot be established from transport measurements alone.

A complementary estimate is provided by the depairing current, which establishes the characteristic scale for homogeneous pair breaking. Following the dirty-limit expression used in Ref. \cite{ref52}, we write $I_{\mathrm{dep}}(0)=\frac{0.74w\Delta_0^{3/2}}{eR_s\sqrt{\hbar D}}$, where $R_s=\rho_N/d$. Using representative NbN parameters $D$ \ensuremath{\approx} 1 cm$^2$/s \cite{ref41}, $d$ = 10 nm, $w$ = 4.5 \ensuremath{\mu}m, and the approximate temperature dependence $I_{\mathrm{dep}}(T)\approx I_{\mathrm{dep}}(0)[1-(T/T_c)^2]^{3/2}$ gives $I_{\mathrm{dep}}(4.5\,\mathrm{K})$ \ensuremath{\approx} 21 mA. The higher measured critical current of approximately 4 mA therefore corresponds to about 19\% of the estimated depairing current. For comparison, the nominal fractions previously estimated for Mo$_2$N/FePt bilayers with Mo$_2$N thicknesses of 10 and 20 nm were approximately 35\% and 45\%, respectively, at their corresponding measurement temperatures \cite{ref23}. Although these estimates depend on representative material parameters and cannot directly quantify differences in vortex pinning between the systems, they show that the experimentally observed dissipation thresholds occur well below the homogeneous depairing limit. This is consistent with vortex-mediated dissipation involving edge injection and potentially the activation of vortices pinned at internal magnetic defects, without uniquely identifying their relative contributions. We next examine how an applied perpendicular magnetic field modifies these polarity-dependent transport regimes and the subsequent flux-flow instability.

\subsection{Perpendicular-field response and flux-flow instability}

\begin{figure*}[tp]
  \centering
  \includegraphics[width=0.91\textwidth,height=0.56\textheight,keepaspectratio]{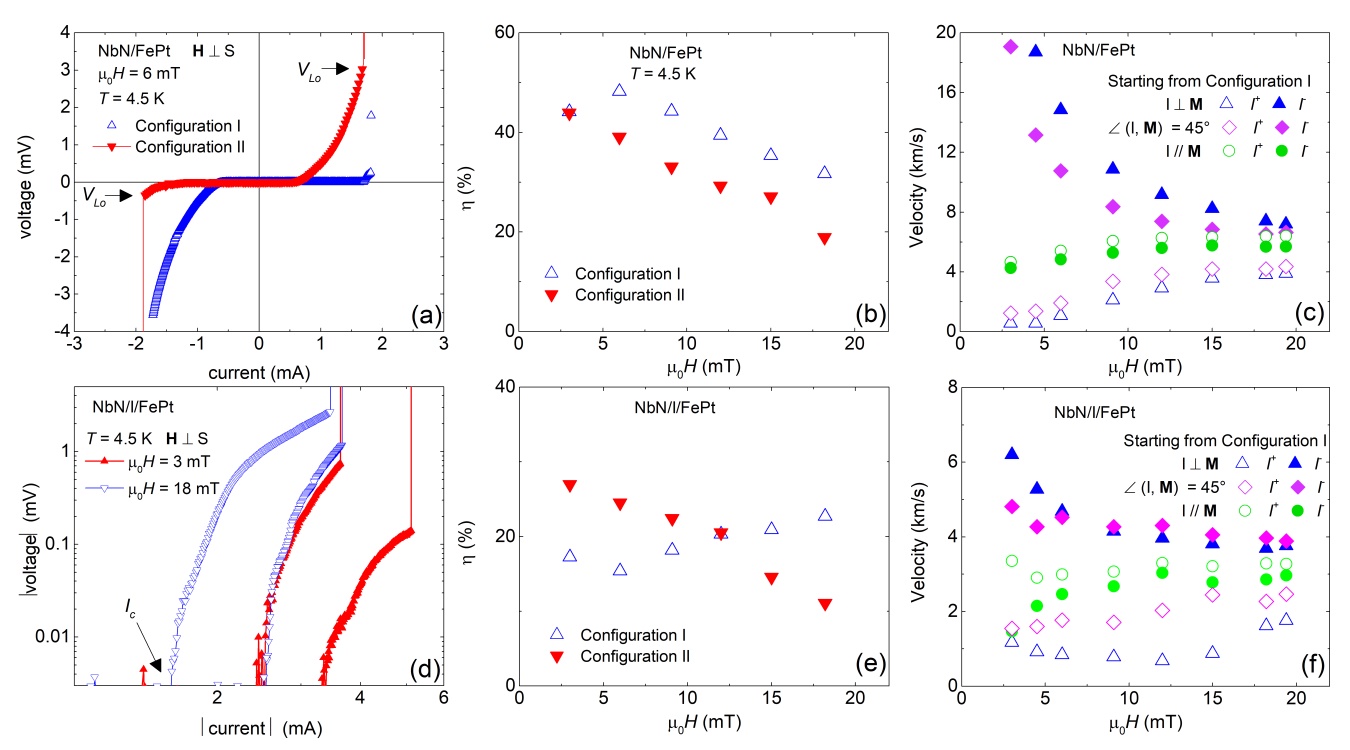}
  \caption{Effect of a perpendicular magnetic field on the superconducting diode effect and apparent instability velocities. (a) I--V characteristics of NbN/FePt at 4.5 K for configurations I and II (stripes perpendicular to the current) under an applied field $\mu_0 H$\ensuremath{\perp} = 6 mT, illustrating the strong dependence of the critical currents and instability voltage $V_{LO}$ on the magnetization configuration. (b) Diode efficiency \ensuremath{\eta} as a function of perpendicular field for NbN/FePt, comparing configurations I and II. (c) Apparent instability velocity as a function of perpendicular field for NbN/FePt, extracted from $V_{LO}$ for both current polarities and different stripe--current orientations, starting from configuration I. (d) |V| vs |I| in log--log scale for selected perpendicular fields in NbN/I/FePt. (e--f) Same as (b--c), respectively, for the NbN/I/FePt sample, highlighting the role of the AlN spacer in the field evolution of the diode efficiency and of the apparent instability velocities.}
  \label{fig:08}
\end{figure*}

Figure 8 examines how an applied perpendicular magnetic field modifies the diode response and the flux-flow instability in the remanent magnetic landscape. The remanent states are prepared by applying and removing an in-plane field before the transport measurements. For the comparison between configurations I and II, the stripe axis is perpendicular to the current (\ensuremath{\theta} = 90\textdegree{}), while the in-plane remanent magnetization is reversed between the two states, as defined in Fig. 5. A static perpendicular field H is then applied during the I--V measurements. Within the proposed picture, this field biases the opposite-signed remanent edge contributions and modifies the conditions for flux injection and subsequent vortex motion. Panel (a) shows representative I--V curves for NbN/FePt at 4.5 K and $\mu_0 H$ = 6 mT. Reversing the remanent magnetization exchanges the branches with lower and higher dissipation thresholds and strongly modifies the polarity dependence of the instability voltage $V_{LO}$. This behavior is consistent with the proposed role of remanent edge fields in controlling polarity-dependent flux injection. Panel (b) shows that the field dependence of the diode efficiency also depends on the remanent configuration. For configuration I, \ensuremath{\eta} reaches nearly 50\% around 6 mT and then decreases, whereas configuration II exhibits a decrease throughout the measured field range. Thus, the applied field tunes the magnitude of the nonreciprocity while preserving a substantial response at the largest fields shown. Panel (c) presents the transport-derived apparent instability velocity, calculated as $v_{\mathrm{LO}}=|V_{\mathrm{LO}}|/(\mu_0HL)$, where $L$ is the voltage-contact separation. The measurements start from configuration I and compare positive and negative current polarities for \ensuremath{\theta} = 90\textdegree{}, 45\textdegree{}, and 0\textdegree{}. For the perpendicular and oblique stripe orientations, the negative-current branch exhibits much larger apparent velocities at low fields, and the contrast between the two polarities decreases as the field increases. The branches are considerably closer when the stripes are parallel to the current. Because the remanent MSF can produce spatially inhomogeneous vortex and antivortex populations, the applied field alone need not characterize the mobile flux responsible for the measured voltage. Accordingly, $v_{LO}$ is treated here as an apparent instability velocity whose interpretation also depends on the effective field on the NbN layers and the mobile-flux distribution.

The corresponding measurements for NbN/I/FePt are shown in panels (d--f) of Fig. 8. Panel (d) displays |V| versus |I| on log--log scales at 4.5 K for $\mu_0 H$ = 3 mT and 18 mT, illustrating the field-induced changes in the dissipation thresholds and instability voltages. Panel (e) reveals distinct field dependences for the two remanent configurations: \ensuremath{\eta} decreases for configuration II, whereas configuration I increases after an initial minimum, with the two efficiencies becoming comparable near 12 mT. Although reversing the remanent magnetization switches the diode polarity, the responses of configurations I and II are not exact mirror images at a fixed perpendicular field. This quantitative asymmetry may reflect differences in the local remanent magnetic texture and vortex-entry barriers at the two bridge edges, including variations between the prepared domain states. Panel (f) again compares the apparent velocities for both current polarities and the three stripe orientations, starting from configuration I. The polarity contrast remains pronounced for perpendicular and oblique stripes and is smaller for parallel stripes. The largest apparent velocities are lower than in NbN/FePt. This difference is compatible with stronger vortex pinning in the less proximity-suppressed NbN layer, while the different reduced temperatures and remanent MSF amplitudes also enter the comparison. The persistence of sizable diode efficiencies up to $\mu_0 H$ = 18 mT shows that the remanently programmed nonreciprocity remains robust within the investigated field range, even as the dissipative branches and instability voltages evolve with the applied field. These millitesla fields are well below the perpendicular saturation field reported for a comparable FePt capping layer \cite{ref40}, supporting an interpretation in which the stripe state is perturbed while its correlated magnetic structure remains operative. Within the proposed picture, the applied field can modify the balance between vortex and antivortex populations, the conditions for flux injection, and the occupation of channels for guided motion, while the remanent edge asymmetry continues to influence the polarity dependence of transport. The evolution of $V_{LO}$ and the associated apparent velocities is therefore consistent with field-sensitive vortex transport within a magnetic landscape that retains the asymmetry responsible for the diode response.

To extend the analysis to higher fields, we measured I--V curves at $T$ = 4.5 K under perpendicular magnetic fields up to 1 T. As mentioned above, \ensuremath{\Lambda} is larger than the bridge width, placing the bridges in the thin-film regime with reduced magnetic screening across their width. Comparing the apparent instability velocities for both stripe orientations across this field range tests how the influence of the initial magnetic configuration evolves with field \cite{ref27,ref28}.

\begin{figure}[tbp]
  \centering
  \includegraphics[width=\columnwidth,height=0.58\textheight,keepaspectratio]{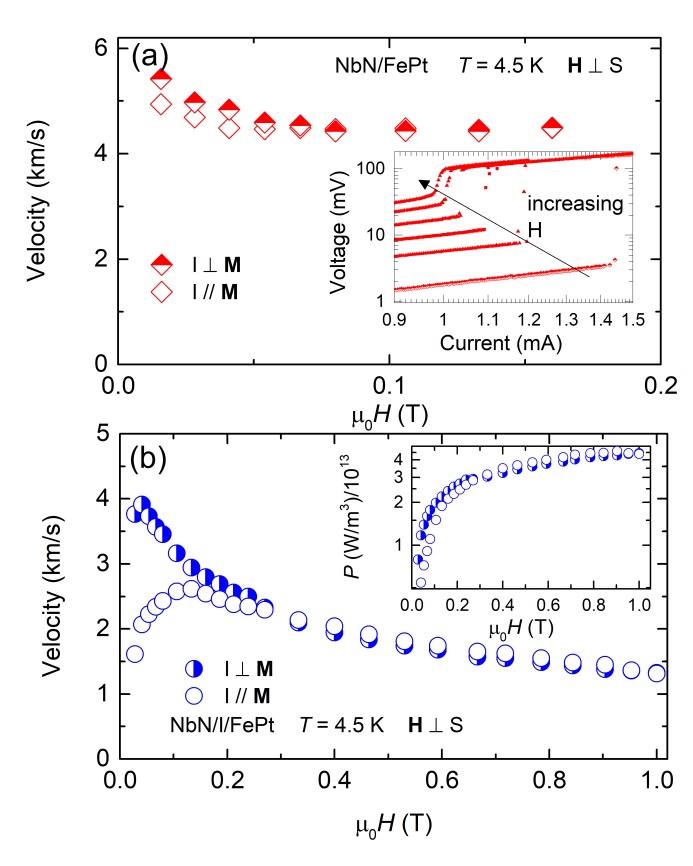}
  \caption{Magnetic-field dependence of the apparent instability velocity at the flux-flow instability for (a) NbN/FePt and (b) NbN/I/FePt at $T$ = 4.5 K with the field applied perpendicular to the film surface (H \ensuremath{\perp} S). Filled (open) symbols correspond to configurations with the magnetization stripes perpendicular (parallel) to the transport current I. The apparent instability velocities were obtained for the positive-current branch (I > 0), starting from remanent configuration II. The inset in (a) shows typical I--V characteristics for NbN/FePt at increasing $\mu_0 H$, with the arrow indicating the increasing field direction, while the inset in (b) displays the power density at the instability as a function of $\mu_0 H$ for both stripe orientations.}
  \label{fig:09}
\end{figure}

Figure 9 presents the magnetic-field dependence of $v_{LO}$ for the positive-current branch I > 0, starting from remanent configuration II, with initial stripe orientations perpendicular and parallel to the current (\ensuremath{\theta} = 90\textdegree{} and 0\textdegree{}, respectively). As shown in Figure 8, the polarity dependence of the apparent instability velocity weakens with increasing applied field. For NbN/FePt [panel (a)], perpendicular stripes yield slightly larger apparent velocities, reaching \ensuremath{\approx} 5.7 km s$^{-1}$ at the lowest fields. The difference between orientations becomes small around 50 mT, and both curves approach approximately 4.5 km s$^{-1}$ over the remaining measured range. In NbN/I/FePt [panel (b)], the orientation dependence is more pronounced and extends to higher fields. The perpendicular configuration exhibits low-field velocities of approximately 4 km s$^{-1}$, whereas the parallel configuration starts around 2 km s$^{-1}$ and initially increases with field. The curves become nearly coincident around 0.2--0.25 T and subsequently decrease together. The different field scales over which the initial stripe orientation influences flux motion may reflect variations in the magnetic domain structure and, as discussed above, in the effective superconducting thickness and associated vortex pinning \cite{ref33,ref34}.

The instability remains identifiable beyond the field at which the orientation contrast weakens. In NbN/FePt, the extracted instability points extend to approximately 0.16 T, whereas in NbN/I/FePt they can be followed up to 1 T. The inset in panel (a) shows representative I--V curves with increasing field, and the inset in panel (b) displays the power density at the instability for both stripe orientations. At higher fields, the power-density curves for both stripe orientations increase toward similar values \cite{ref36}, tending toward approximately 4 \ensuremath{\times} 10$^{13}$ W m$^{-3}$. The plateau in $v_{LO}$ ($H$) for NbN/FePt is compatible with the homogeneous LO limit, in which the intervortex spacing a is small compared with the quasiparticle diffusion length $\ell_e=\sqrt{D\tau}$ \cite{ref8}. This field dependence alone does not establish a purely non-thermal LO mechanism. Unlike the enhancement relative to uncapped Mo$_2$N reported in Refs. \cite{ref27,ref40}, the instability velocities in the present devices do not exceed those reported for uncapped NbN films \cite{ref37,ref53}. We refrain from extracting \ensuremath{\tau} because of uncertainties in the effective induction and the appropriate transport parameters for the proximity-affected NbN layer \cite{ref35}.

The evolution of $v_{LO}$ ($H$) with field reflects both the superconducting properties of the effective NbN layer and the progressively weaker influence of the initial stripe orientation on vortex motion. The larger and more persistent orientation contrast in NbN/I/FePt may reflect a greater relative benefit of stripe-guided motion if pinning is stronger in the less proximity-suppressed layer. In NbN/FePt, the reduced effective superconducting thickness and weaker pinning may account for the higher velocities and smaller contrast between orientations. These differences, together with those in $T_c$, support the role of interfacial transparency in modifying the superconducting state and dissipative dynamics beyond the contribution of the remanent MSF \cite{ref11,ref37,ref38}.

\section{Summary and conclusions}

We have demonstrated reconfigurable, field-free superconducting diode behavior and orientation-dependent vortex dynamics in NbN/FePt and NbN/AlN/FePt heterostructures based on epitaxial NbN films approximately 10 nm thick. The FePt layers develop stripe domains approximately 50 nm wide with a period of approximately 100 nm, whose orientation can be set in remanence. Introducing a thin AlN spacer retains a comparable NbN(111) peak width and raises $T_c$ from approximately 9.8 to 15.5 K, consistent with reduced electronic proximity coupling while retaining magnetic coupling through stray fields.

For stripes perpendicular to the current, the diode efficiency reaches approximately 40-- 45\% in NbN/FePt and 20--25\% in NbN/AlN/FePt at 4.5 K. Both efficiencies increase overall with temperature, reaching approximately 60\% and 35--40\%, respectively, at the highest measured temperatures. Rotating the stripes toward alignment with the current reduces the nonreciprocity. Its persistence with the insulating spacer supports a magnetic interpretation in which opposite-signed remanent edge fields favor polarity-dependent flux injection, while the alternating field of neighboring stripes provides confinement and channels for guided vortex motion. Both current polarities develop finite voltage before switching, departing from the ideal edge-injection regime. The substantial dissipation threshold of the lower-current branch is consistent with significant pinning that limits propagation along the stripe channels in addition to the barriers governing flux injection. Meanwhile, bifurcations and internal terminations of the magnetic stripes introduce localized stray-field variations that may favor vortex nucleation and strong magnetic confinement within the bridge. Current-driven activation of this internal population provides a plausible contribution to dissipation in the higher-current branch, even when injection of the preferential edge vorticity is suppressed. The results therefore support a distinction between the edge asymmetry that selects the diode polarity and the internal pinning landscape that controls the onset and development of dissipation. Transport alone cannot distinguish a vortex-depleted interior from one containing immobilized vortices below the voltage criterion.

Applied perpendicular fields modify the diode efficiency and the instability voltage $V_{LO}$ in a manner that depends on the remanent configuration. Reversing the remanent magnetization exchanges the branches with lower and higher dissipation thresholds, while sizable nonreciprocity persists throughout the investigated millitesla field range, up to 18 mT. At larger fields, the contrast between the apparent instability velocities for perpendicular and parallel stripes becomes small around 50 mT in NbN/FePt and 0.2--0.25 T in NbN/AlN/FePt, while the instability remains measurable beyond these fields. Despite its lower apparent velocities, the sample with the spacer exhibits a larger and more persistent orientation contrast, which may reflect a greater relative benefit of stripe-guided motion if intrinsic pinning is stronger.

Using NbN/FePt bilayers and NbN/AlN/FePt heterostructures, we extend the remanent magnetic control previously demonstrated in Mo$_2$N/FePt \cite{ref23} to higher operating temperatures. The diode response retains sizable nonreciprocity under weak perpendicular fields. Although lower than in Mo$_2$N/FePt, the maximum zero-field efficiency of approximately 60\% exceeds the approximately 40\% reported for NbN/AlN/Py heterostructures \cite{ref18} and is comparable to the 54\% reported for geometrically asymmetric Pt/Co/Nb devices \cite{ref17} and to the 65\% obtained in Pt/V/EuS under a small applied in-plane field by Hou $et$ $al.$ \cite{ref10}. Beyond the efficiency comparison, NbN/AlN/Py also exhibits dissipation before switching in both current polarities \cite{ref18}. Current-driven activation of initially pinned interior vortices could contribute to this shared behavior, particularly in the high-current branch, while asymmetric flux injection and expulsion at the edges can continue to select the diode polarity. This interpretation suggests magnetic-domain regularity and longitudinal vortex pinning as complementary design parameters for stripe-controlled superconducting diodes. Reducing internal domain defects could suppress defect-assisted nucleation and premature dissipation in the high-current branch, while weak pinning along regular stripe channels could lower the threshold of the opposite branch. Provided that the asymmetric edge-injection conditions are preserved and the regular interior remains depleted of mobile vortices for the high-current polarity, this combination offers a route toward simultaneously increasing the high-current threshold toward the depairing scale and reducing the low-current threshold. Approaching this limit requires joint control of the magnetic texture, superconducting properties, and vortex-flow stability.

\section*{Author information}

Corresponding Authors: J. Kim (jeehoon@postech.ac.kr); N. Haberkorn (nestor.haberkorn@ib.edu.ar).

\section*{Author contributions}

N. Haberkorn and J. Kim conceived the project. E. De Biasi performed the micromagnetic simulations. N. Haberkorn and J. Zapata prepared the samples and carried out XRD and electrical transport measurements. M. Sirena acquired AFM and MFM images. All authors discussed the results and contributed to writing and revising the manuscript. All authors have approved the final version of the manuscript.

\section*{Acknowledgements}

The authors acknowledge S. Anguiano for technical assistance. This work was partially supported by the CONICET (PIP 11220210100263CO), BrainLink program funded by the Ministry of Science and ICT through the National Research Foundation of Korea (No. RS-2024-00410027 and No. RS-2026-25472018) and Brain Pool program funded by the Ministry of Science and ICT through the National Research Foundation of Korea (RS-2023-00222408). Computational resources were provided by the HPC cluster of the Physics Department at Centro Atómico Bariloche (CNEA). We acknowledge the Texas Advanced Computing Center (TACC) for providing the computational hardware and the Physics Department at Centro Atómico Bariloche (CNEA) for hosting and operating the HPC cluster used in this work.


\end{document}